\documentclass[journal,twoside,web]{ieeecolor}
\usepackage{lcsys}
\usepackage{cite}
\usepackage{amsmath,amssymb,amsfonts}
\usepackage{algorithmic}
\usepackage{subfigure}
\usepackage{graphicx}
\usepackage{textcomp}
\newcommand{\lb}{\left(}
\newcommand{\rb}{\right)}
\newcommand{\ls}{\left[}
\newcommand{\rs}{\right]}

\newcommand{\bs}[1]{\boldsymbol{#1}}
\newcommand{\cl}[1]{\mathcal{#1}}
\newcommand{\bb}[1]{\mathbb{#1}}
\def\BibTeX{{\rm B\kern-.05em{\sc i\kern-.025em b}\kern-.08em
    T\kern-.1667em\lower.7ex\hbox{E}\kern-.125emX}}
\begin{document}
\title{Networked Control System Under Controller-Actuator Channel Jamming}
\author{Chen Quan, Geethu Joseph, and Gourab Ghatak
\thanks{C. Quan and G. Joseph are with the Faculty of Electrical Engineering, Mathematics, and Computer Science,
        TU Delft, 2628 CD Delft, Netherlands. Emails: {\tt\small\{c.quan,g.joseph\}@tudelft.nl.}}
\thanks{G. Ghatak is with the Department of Electrical Engineering, IIT Delhi, Delhi 110016 India. Email: {\tt\small gghatak@ee.iitd.ac.in. }}
}

\maketitle

\begin{abstract}
Wireless channels in the networked control systems are vulnerable to intentional interference, such as jamming attacks. This paper investigates jamming attacks on the wireless controller-actuator channel of a control system that can tolerate occasional control inputs from the controller. We start with a worst-case scenario for the jammer where the controller knows its channel state. We develop an adaptive jamming strategy in which the jammer, observing the success or failure of each controller transmission, forms beliefs about its own and the controller–actuator channel states. Using this belief, it optimizes its actions under a limited jamming budget.  To counter this, we develop an event-triggered defense scheme for the controller in two settings: with and without the knowledge of its channel state. Simulation results show that optimal adaptive jamming attacks can significantly degrade control performance, even with a limited budget, while the defense scheme, even without channel state knowledge, can effectively reduce this impact.
\end{abstract}


\section{Introduction}
\label{sec:introduction}
\IEEEPARstart{N}{etworked} control systems, used in autonomous vehicles, smart grids, industrial automation, and remote robotics, rely on wireless communication to coordinate sensing, control, and actuation. While wireless communication offers flexibility and scalability, it is vulnerable to intentional interference, such as Byzantine~\cite{lamport2019byzantine} and denial-of-service attacks~\cite{raymond2008denial}. In this paper, we focus on jamming interference~\cite{d2016optimal}, one of the most threatening denial-of-service attacks.
 
Existing studies on jamming examine different attacking schemes and tailored defense strategies, focusing its impact on control performance and stability~\cite{7172466,7445829,6425868,8673781,shisheh2016triggering,7070677,senejohnny2017jamming,
9133451,10938979,9419856,9832494,8561209}.
For example, game-theoretic methods ensure stability under periodic jamming on sensor-controller channels~\cite{7172466,7445829}, event-triggered controls stabilize plants under jamming on controller–actuator channels~\cite{6425868,8673781}, and combined channel attacks are handled via event-triggered control and optimized transmission schedules~\cite{shisheh2016triggering,7070677,senejohnny2017jamming}. 
Beyond periodic jamming, aperiodic jamming with irregular timing and duration~\cite{9133451,10938979,9419856}, and random jamming with probabilistic attack patterns~\cite{9832494,8561209} have been studied.
However, these works overlook intelligent jammers that quickly adjust strategies using real-time feedback. Moreover, stability is not the only desired property of controlled systems; in many applications, such as underactuated mechanical systems~\cite{huang2024survey} and switched systems~\cite{klamka2013controllability}, controllability is a more fundamental requirement.
We examine how adaptive aperiodic jamming on the controller-actuator channel impacts the controllability of systems that can tolerate occasional communication failures.

Our contributions are two-fold: First, we develop an optimal strategy for the jammer that dynamically adjusts its attack based on a joint belief about its own and the controller's channel, under a constrained budget. Second, to defend such attack, we present an event-triggered defense scheme for the controller and actuator, which are initially unaware of the jammer, using its belief about the controller-actuator channel state. This design enhances system resilience while reducing communication overhead in the system. Simulation results show the defense scheme improves controllability and effectively counters adaptive jamming attacks.

\section{Control System and Jamming Model}
Consider a networked control system with a remote controller that communicates with the sensor and actuators through wireless channels. The system state $\bs{x}(t)\in\bb{R}^n$ evolves according to a discrete-time linear dynamical system, 
\begin{equation}\label{eq:controlsys_model}
    \bs{x}(t+1) = \bs{A}\bs{x}(t)+\bs{B}\bs{u}(t)+\bs{v}(t),
\end{equation}
where $\bs{A}\in\bb{R}^{n\times n}$ and $\bs{B}\in\bb{R}^{n\times m}$ denote the state and input matrices. Also, $\bs{u}(t)\in\bb{R}^m$ and $\bs{v}(t)\in\bb{R}^n$ denote the control input and zero-mean Gaussian process noise independent of $\bs{x}(t)$ and $\bs{u}(t)$, respectively, at discrete time $t\in\bb{Z}_+$. 
The system aims to drive and maintain $\bs{x}(t)$ at a desired state $\bs{x}_{\rm des}$ via appropriate control inputs~$\bs{u}(t)$.

We assume the setting in~\cite{10778310,11202619} where the actuator has limited computing capabilities and all computations are offloaded to the remote controller. The controller receives system state $\bs{x}(t)$  from the sensor at the beginning of every block of $T$ time steps, i.e., at times $t=kT$, where $k\in\bb{Z}_+$ is the block index. For simplicity, we assume that the sensor-controller channel is reliable due to its sporadic usage, implying the controller knows the true state $\bs{x}(kT)$.  
Then, the controller computes the control inputs to drive the state to $\bs{x}_{\rm des}$ and transmits them to the actuator in sequential batches.

\begin{figure}[hpt!]
        \centering
        \includegraphics[width = 7cm, height=4cm]{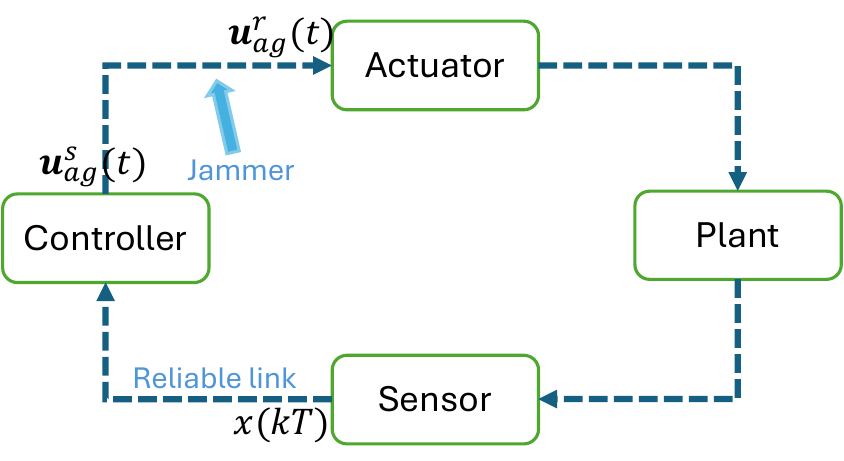}
        \caption{Control system under jamming:  At the start of block $k$, the sensor sends state $\mathbf{x}(kT)$ to the controller via a reliable link. 
        The controller sends aggregated control inputs $\bs{u}^s_{\mathrm{ag}}(t)$ in batches over time slots within block $k$ to the actuator,
        disrupted by a jammer, resulting in received inputs~$\bs{u}^r_{\mathrm{ag}}(t)$.}
        \label{fig:sys}
\end{figure}

In this paper, we consider the presence of an adaptive jammer in the network, as shown in Fig.~\ref{fig:sys}. The jammer interferes with the controller-actuator channel, so the control inputs may not always be successfully decoded by the actuator. The actuator broadcasts the transmission acknowledgment (whether the control input was correctly received) to both the controller and jammer. They use the feedback to adapt their strategies, the jammer to hinder, and the controller to enhance control performance. 


\subsection{Controller Channel Model Under Jamming Attacks}\label{sec:channelmodel}
We denote the $\bs{u}^s_{\mathrm{ag}}(t)$ as the aggregated control inputs transmitted from the controller to the actuator at time slot $t$. The received signal at the actuator affected by both jammer interference and channel noise is
\begin{equation}\label{eq:receivedcontrol}
    \bs{u}^r_{\mathrm{ag}}(t) = h_{\mathrm{Tx}}(t) \bs{u}^s_{\mathrm{ag}}(t) + h_{\mathrm{Jm}}(t)a_t\bs{w}_{\mathrm{Jm}}(t)+\bs{w}_{\mathrm{Tx}}(t),
\end{equation}
where $h_{\mathrm{Tx}}(t)$ and $h_{\mathrm{Jm}}(t)$ are the fading coefficients for the controller-actuator and the jammer-actuator channels, respectively. Also, $a_t\in\{0,1\}$ is the jammer's action, which indicates whether the jammer is active or not at time slot $t$. The terms $\bs{w}_{\mathrm{Tx}}(t)$ and $\bs{w}_{\mathrm{Jm}}(t)$ are zero-mean Gaussian random vectors, denoting the inherent channel noise and the jamming signal at time slot $t$, respectively. 
We assume $\bs{w}_{\mathrm{Tx}}(t) \sim \mathcal{N}(0, \sigma_{\mathrm{Tx}}^2 \bs{I})$ and $\bs{w}_{\mathrm{Jm}}(t) \sim \mathcal{N}(0, \sigma_{\mathrm{Jm}}^2 \bs{I})$, where $\sigma_{\mathrm{Tx}}^2$  and $\sigma_{\mathrm{Jm}}^2$ are the noise variances.

\subsubsection{Channel Model}
We model both the controller-actuator and jammer-actuator channels as fast Rayleigh fading with block-wise stationarity. Specifically, within each block, the Rayleigh parameters remain constant and are determined by the current channel state. Each channel can be in one of two states, $G$ (good) or $B$ (poor), characterized by Rayleigh parameters $\theta_G$ and $\theta_B$ with $\theta_B > \theta_G$. The controller and jammer channel states in block $k$, denoted by $s_k^{\mathrm{Tx}}$ and $s_k^{\mathrm{Jm}}$, follow independent Markov chains with states $\{G, B\}$. The Markov chains are defined by prior probabilities $\pi_q^{\mathrm{Tx}}$ for $s_k^{\mathrm{Tx}} = q$ and $\pi_q^{\mathrm{Jm}}$ for $s_k^{\mathrm{Jm}} = q$, and by transition probabilities $p_{q,d}^{\mathrm{Tx}}$ and $p_{q,d}^{\mathrm{Jm}}$ from state $q$ to $d$ for the controller and jammer, respectively, where $q, d \in \{G, B\}$. The joint channel state is $s_k = (s_k^{\mathrm{Tx}}, s_k^{\mathrm{Jm}})$, and the corresponding Rayleigh parameters are $\theta^{\mathrm{Tx}}(k) = \theta_q$ and $\theta^{\mathrm{Jm}}(k) = \theta_d$ if $s_k = (q, d)$.

\subsubsection{Transmission Success}
Due to the stochastic noise and jamming, the actuator may not always successfully decode the control inputs from the controller. The transmission at time slot $t$ is considered successful, indicated by $S_t
\in\{0,1\}$, if the received SNR exceeds a given threshold $\gamma>0$, determined by the application and receiver hardware,
\begin{equation*}
    \mathrm{SNR}(t) = \frac{|h_{\mathrm{Tx}}(t)|^2 \eta}{\sigma_{\mathrm{Tx}}^2 + |h_{\mathrm{Jm}}(t)|^2 a_t\sigma^{2}_{\mathrm{Jm}}}\geq\gamma,
\end{equation*}
where $\eta$ is the controller transmit power.
The success probability in time slot $t$ of block $k$, conditioned on the controller and jammer channels and the jammer's action, is
\begin{equation}\label{eq:successprob}
    P_{q,d,a} = \mathrm{Pr}\left(S_t=1\mid s_k=(q,d), a_t=a\right).
\end{equation}
When the jammer is active, $a_t=1$, we deduce
\begin{equation*}
    P_{q,d,1}=\mathrm{Pr}\left(|h_{\mathrm{Tx}}(t)|^2 \eta > \gamma\sigma^{2}_{\mathrm{Tx}} + \gamma |h_{\mathrm{Jm}}(t)|^2 \sigma^{2}_{\mathrm{Jm}}\right).
\end{equation*}
Since squared Rayleigh variables are exponentially distributed, $|h_{\mathrm{Tx}}(t)|^2\sim\exp(\theta_q)$ and $|h_{\mathrm{Jm}}(t)|^2\sim\exp(\theta_d)$, and
    \begin{equation*}
        P_{q,d,1}
    =\bb{E}_{h_{\mathrm{Jm}}(t)}\ls e^{-\theta_q \frac{\gamma\sigma^{2}_{\mathrm{Tx}} + \gamma |h_{\mathrm{Jm}}(t)|^2 \sigma^{2}_{\mathrm{Jm}}}{\eta}}\rs=\frac{\theta_d e^{-\frac{\theta_q \gamma \sigma^{2}_{\mathrm{Jm}}}{\eta}}}{\theta_q \frac{\gamma \sigma^{2}_{\mathrm{Jm}}}{\eta} + \theta_d}.
    \end{equation*}
With an inactive jammer, the success is independent of the jammer channel, and for $d\in\{G,B\}$, 
\begin{equation}\label{eq:successprob2}
        P_{q,d,0}=P_{q,0}
        =\mathrm{Pr}\left(|h_{\mathrm{Tx}}(t)|^2 > \frac{\gamma\sigma^{2}_{\mathrm{Tx}}}{\eta} \right)=e^{-\frac{\theta_q \gamma \sigma^{2}_{\mathrm{Tx}}}{\eta}}.
    \end{equation}
Based on the above success probabilities, the jammer aims to degrade control performance with minimal jamming activities to save energy. The control performance depends on the overall controller-actuator design, described next.

\subsection{Control System Model}
We adopt the restless system model in~\cite{10778310,11202619} to describe the control system, as elaborated below.
\subsubsection{Control Input Design at Controller}
In each block $k$, the controller designs the control input based on the received state $\bs{x}(kT)$. Since the noise $\bs{v}(t)$ is zero-mean and unknown to the controller, the state estimate at time $t>kT$ is
\begin{equation}\label{eq:estimate}
    \hat{\bs{x}}(t) = \bs{A}^{t-kT}\bs{x}(kT)+\sum_{\tau=kT}^{t-1}\bs{A}^{t-\tau-1} \bs{B}\bs{u}(\tau).
\end{equation}
The controller designs $v<T$ control inputs $\{\hat{\bs{u}}(t+\tau),\tau=0,1,\ldots,v-1\}$ such that $\hat{\bs{x}}(t+v)=\bs{x}_{\rm des}$. 
To ensure controllability, we choose $v$ as the controllability index of the dynamical system $(\bs{A},\bs{B})$ in \eqref{eq:controlsys_model}, which is assumed to be known to the controller, actuator, and jammer. The control inputs are obtained via least squares~\cite{10778310}
\begin{multline*}\label{eq:controldesign_est}
    \begin{bmatrix}
        \hat{\bs{u}}(t)^{\top}&
        \hat{\bs{u}}(t+1)^{\top}
        \ldots
        \hat{\bs{u}}(t+v-1)^{\top}
    \end{bmatrix}^{\top} \\= \begin{bmatrix}
    \bs{A}^{v-1}\bs{B} & \bs{A}^{v-2}\bs{B} & \ldots & \bs{B}
\end{bmatrix}^{\dagger}\ls \bs{x}_{\rm des} - \bs{A}^{v}\bs{x}(kT)\rs.
\end{multline*}

\subsubsection{Control Input Transmission}
The controller adapts the number of control inputs per slot based on channel quality.  For now, we assume the controller knows its channel state $s_k^{\mathrm{Tx}}$, which will be relaxed in Section~\ref{sec:unknownchannel}. For the good state ($s_k^{\mathrm{Tx}}=G$), it uses higher-order modulation to transmit $\mu(k)=\mu_G$ control inputs per slot. In the poor state ($s_k^{\mathrm{Tx}}=B$), it switches to lower-order modulation for reliability, sending only $\mu(k)=\mu_B < \mu_G$ inputs. Then, at time slot $t$ in block $k$, the controller sends an aggregated control input 
\begin{equation*}
    \bs{u}_{\mathrm{ag}}^s(t) = \begin{bmatrix}
\hat{\bs{u}}^{s}(t)&\hat{\bs{u}}^{s}(t+1)&\dots&
\hat{\bs{u}}^{s}(t +\mu(k) - 1)
\end{bmatrix}^{\top},
\end{equation*}
where $\hat{\bs{u}}^{s}(t)=[\hat{\bs{u}}(kT+\mu(k)L(t))]^{\top}$ and $L(t)=\sum_{\tau=kT}^{t-1}S_\tau$ denotes the number of successful transmissions up to time $t-1$. After transmission at time $t$, the controller receives an acknowledgment $S_t$. If $S_t = 0$, the controller retransmits the previously failed inputs; if $S_t=1$, it proceeds to send the next set of control inputs until all $v$ inputs have been successfully received, i.e., until $\mu(k)L(t)\geq v$.

\subsubsection{Actuator-Side Control Input}
The actuator receives the noisy aggregated control inputs $\bs{u}_{ag}^r(t)\in\bb{R}^{m \mu(k)}$, given in \eqref{eq:receivedcontrol}. If the SNR exceeds $\gamma$ (i.e., $S_t=1$), it correctly decodes and applies the control inputs from the controller to the plant. If the transmission fails and the actuator has not yet received the necessary control input from the controller, it switches to a feedback mechanism to compensate for the missing inputs. Since restless system assumption~\cite{11202619} requires the column space of $\bs{I} - \bs{A}$ lies within that of $\bs{B}$, to ensure that when $S_t=0$, the system can maintain $\hat{\bs{x}}(t+1)=\hat{\bs{x}}(t)$. This is achieved through the state-dependent feedback $ \bs{u}(t)=\bs{B}^{\dagger}\lb \bs{I} - \bs{A}\rb\bs{x}(t)$, without the need to recompute the entire control input.  
Once all $v$ control inputs have been applied and $\hat{\bs{x}}(t) = \bs{x}_{\rm des}$, the system switches to feedback-based control for the remainder of the block. So, $\hat{\bs{x}}(t)=\bs{x}_{\rm des}$ until the next block.

Under this model, control performance is measured by block controllability~\cite{11202619}, achieved when the actuator drives the estimate to the desired state, i.e., $\hat{\bs{x}}(t) = \bs{x}_{\rm des}$, or equivalently when $\mu(k)L(t)\geq v$.
Next, we discuss the optimal jamming strategy and the defense scheme.

\section{Optimal Jamming Strategy and Defense}
We begin with the jammer’s goal: minimizing the success of the controller's transmission with minimal jamming. To this end, the jammer maintains a posterior belief over the four possible channel-state combinations based on transmission acknowledgments. It then selects actions that degrade block controllability while conserving energy, assuming a worst-case scenario where the controller knows its own channel state. To counter such intelligent and adaptive jamming, we then present a defense strategy from the controller’s side.
        
\subsection{Belief Updating Model}
We denote the belief model of the jammer at time slot $t$ in block $k$ as 
\begin{equation}\label{eq:b_kqdt}
    b_{k,q,d}(t)=\mathrm{Pr}(s_k=(q,d)\mid \{S_{\tau},a_{\tau}\}_{\tau=kT}^{t-1}),
\end{equation}
for $q,d \in \{G,B\}$, computed based on all the acknowledgments and jamming actions up to time $t-1$ in block $k$.
Given $S_t$ and $a_t$ at time slot $t$, the new belief at time slot $t+1$ is 
\begin{equation}\label{eq:joint_update}
    b_{k,q,d}(t+1)= \frac{b_{k,q,d}(t)P_{q,d,a_t}
   }{
      \sum_{q',d'} \!b_{k,q',d'}(t)P_{k,q',d',a_t}},
\end{equation}
using \eqref{eq:successprob}. The update equations use the initial joint belief given by $\pi^{\mathrm{Tx}}_q\pi^{\mathrm{Jm}}_d$, where $\pi^{\mathrm{Tx}}_q$ and $\pi^{\mathrm{Jm}}_d$ are the prior probabilities of the channel states.
From \eqref{eq:joint_update}, the marginal belief of the controller 
channel is $b^{\mathrm{Tx}}_{k,q}(t)=b_{k,q,G}(t)+b_{k,q,B}(t)$. 

Next, we formulate optimization problems from the jammer’s perspective, for single-block and multi-block scenarios, to prevent block controllability while minimizing jamming.

\subsection{Optimal Jamming: Single-block Scenario}
In the single-block scenario, the jammer determines its jamming actions probabilistically based on the number of remaining time slots within the current block and past transmission acknowledgments from the actuator. Specifically, at time slot $t$, it computes an optimal jamming probability $g_{t,k}^*$, and the jamming action is drawn as $a_t\sim \text{Bernoulli}(g_{t,k}^*)$.

To build its belief, the jammer listens passively in the first time slot ($t=kT$) of every block, observing the controller's channel to plan future actions. At time slot $t>kT$ in block $k$, the jammer minimizes both the expected total number of jamming events and the average number
of successful transmissions in the remaining $\Delta T=(k+1)T-t$ time slots of the block, 
\begin{multline}
    e_s(g_{t,k};t)=\sum_{q,d}\!\mu_{q}P_{q,d,1}b_{k,q,d}(t)  \Delta Tg_{t,k}\\
    +\sum_{q}\mu_{q}P_{q,d,0}b_{k,q}^{\mathrm{Tx}}(t)\left\{\Delta T(1-g_{t,k})\right\}.\label{eq:es}
\end{multline}
So, the jammer's optimization problem is
\begin{multline*}
    \min_{ g_{t,k}\in[0,1] } 
   \;\;e_s(g_{t,k};t)+\lambda \Delta T g_{t,k}
   \quad
   \\\text{s.t.}\; \textstyle\bb{E}\ls\sum_{\tau=t}^{(k+1)T-1}S_{\tau}\rs\!\leq\! \sum_{q\in\{G,B\}} b^{\mathrm{Tx}}_{k,q}(t)\frac{v}{\mu_q}-
   \sum_{\tau=kT}^{t-1}S_{\tau},
\end{multline*} 
where $\lambda$ is a penalty coefficient on jamming actions.
The constraint ensures that the number of future successful transmissions in the block, $\bb{E}[\sum_{\tau=t}^{(k+1)T-1}S_{\tau}]$, does not exceed the average number required for controllability. We simplify
\begin{multline}
    \textstyle\bb{E}\ls\sum_{\tau=t}^{(k+1)T-1}S_{\tau}\rs=\sum_{q,d}P_{q,d,1}b_{k,q,d}(t)\Delta Tg_{t,k}\\+\textstyle\sum_{q}P_{q,d,0}b^{\mathrm{Tx}}_{kq}(t)\{\Delta T(1-g_{t,k})\}.\label{eq:define_p_nqdl}
\end{multline}
Using \eqref{eq:es} and \eqref{eq:define_p_nqdl}, the jammer's optimization reduces to the following convex optimization formulation:
\begin{equation}\label{eq:optimization_simple}
    \min_{ g_{t,k}\in[0,1] } 
   \;\;(\alpha_t+\lambda) g_{t,k}
 \quad\text{s.t.}\quad \Delta T\rho_tg_{t,k}\leq R_t,
\end{equation}
where we define
\begin{align*}
    \alpha_t &= \sum_{q,d}\mu_{q}P_{q,d,1}b_{k,q,d}(t)  -
    \sum_{q}\mu_{q}P_{q,d,0}b_{k,q}^{\mathrm{Tx}}(t)\\
    \rho_t & = \sum_{q,d}P_{q,d,1}b_{k,q,d}(t)-\sum_{q}P_{q,d,0}b^{\mathrm{Tx}}_{k,q}(t)\\
    R_t&=
    \sum_{q\in\{G,B\}} b^{\mathrm{Tx}}_{k,q}(t)\frac{v}{\mu_q}\!-\!
   \sum_{\tau=kT}^{t-1}S_{\tau}
   -\Delta T\sum_{q}P_{q,d,0}b^{\mathrm{Tx}}_{k,q}(t).
\end{align*}
Moreover, \eqref{eq:optimization_simple}, if feasible, admits a closed-form solution as
\begin{equation*}
     g_{t,k}^* \!=\!
\begin{cases}
0& \text{if } \alpha_t \geq -\lambda, \rho_t\geq 0 \\
\max(\min\left\{ R_t/(\rho_t\Delta T) , 1 \right\},0) & \text{if } \alpha_t \geq -\lambda, \rho_t<0 \\
\max(\min\left\{  R_t/(\rho_t\Delta T), 1 \right\},0) & \text{if } \alpha_t \!<\! -\lambda, \rho_t>0\\
1 & \text{if } \alpha_t\!<\! -\lambda, \rho_t\leq 0.
\end{cases}
\end{equation*}
If the optimization problem in \eqref{eq:optimization_simple} is infeasible, the jammer chooses $g_{t,k}^*=0$ to save energy. 


\subsection{Optimal Jamming: Multi-block Scenario}
In the multi-block scenario, the jammer optimizes its jamming probabilities over $K$ successive blocks under a limited jamming budget. Here, the jammer updates the belief at the beginning of each block using the posterior from the previous block using the Markov chain's transition probability. For $q,d\in\{G,B\}$, the belief at the start of block $k$ is
\begin{equation*}
b_{k,q,d}(kT)= \sum_{q,d\in\{G,B\}} 
b_{k-1,q',d'} (kT-1)
p_{q',q}^{\mathrm{\mathrm{Tx}}}\, p_{d',d}^{\mathrm{\mathrm{Jm}}}.
\end{equation*}
Similar to \eqref{eq:optimization_simple}, the optimization problem at time slot $t$ in block $k$ under a total jamming budget of $M$ is given by
\begin{multline}\label{eq:opt_2}
    \min_{ \substack{g_{t,k}\in[0,1]\\ \{g_{k'}\in[0,1]\}_{k'=k+1}^K}} \Delta T(\alpha_{t}+\lambda_{\text{d}})g_{t,k}+T\sum_{k'=k+1}^K (\alpha_{k'T}+\lambda_{\text{d}})g_{k'} \\
    \quad\text{s.t.}\quad \Delta T\rho_tg_{t,k}\leq R_t, \quad \rho_{k'T}g_{k'}\leq R_{k'T},  \text{ for } k'>K\\
    \Delta Tg_{t,k}+T\sum_{k'=k+1}^{K-1} g_{k'}\leq M-\sum_{t'=0}^{t-1} a_{t'},
\end{multline}
where $g_{k'}=g_{k'T,k}$ is the jamming probability applied at each time slot within block $k'$, and $M$ is the total number of time slots the jammer can jam over the $K$ blocks. The term $\lambda_{\text{d}}$ is a dynamically changing penalty coefficient on jamming actions based on belief updating at time slot $t$, which is given by $\lambda_{\text{d}}=\sum_{q,d}\lambda_{q,d}b_{k,q,d}$, where $\lambda_{q,d}$ is the penalty coefficient associated with the the controller and jammer channel states $(q, d)$. In the last constraint in \eqref{eq:opt_2}, the term $\sum_{t'=0}^{t} a_{t'}$ is the total number of past jamming events till time $t-1$, ensuring that the total number of jamming actions across all blocks does not exceed the jammer’s budget $M$. This convex optimization problem is solved using standard toolboxes. If constraint $\Delta T\rho_tg_{t,k}\leq R_t$ makes the problem infeasible, we set $g_{t,k}=0$, and proceed to solve for $g_{k'}$ for $k' \in \{k+1, \dots, K-1\}$. However, if the final budget constraint renders the problem infeasible, we set $g_{t,k}=g_{k'}=0$ for $k'\geq k+1$. 
  
Setting $K=1$ in \eqref{eq:opt_2} gives a variant of single-block optimization with a budget constraint. Further, setting $K=1$, $M=T$, and $\lambda_{q,d}=\lambda$ in \eqref{eq:opt_2} leads to optimization in~\eqref{eq:optimization_simple}.  

\subsection{Defense Strategy} 
For defense, we adopt a `deceiving the jammer' strategy, triggered under one of two events: $\cl{E}_{\rm{bad}}$ when jammer activity is detected in a block with poor controller channel, or $\cl{E}_{\rm{insuff}}$ when the remaining slots are insufficient to ensure controllability. Then, the controller sends a one-bit command to the actuator via a reliable, jammer-proof channel, instructing the actuator to broadcast false success acknowledgments. Unlike control signal transmission, the command has a small data size, allowing strong forward error correction to ensure reliable reception of this bit. 
These false acknowledgments can mislead the jammer into believing its jamming is ineffective, encouraging it to waste the jamming budget in the current block. Particularly, if $s_k=(B,G)$, false feedback can cause the jammer to allocate more budget on the current block, reducing jamming in future blocks.

\subsubsection{Controller channel state known at the controller}
The controller does not know the jammer’s presence or actions, so it first estimates the likelihood of jamming. Let $z_k\in\{0,1\}$ indicate the presence of jamming in block $k$, and 
\begin{equation*}
    f_{k}^{\mathrm{Jm}}(t)=\mathrm{Pr}(z_k=1|\{S_\tau\}_{\tau=kT}^{t-1})\in [0,1],
\end{equation*}
be the belief, with an initial value of $f^{\mathrm{Jm}}_{k}(kT)=0.5$. 

The belief is updated based on the difference between the observed transmission success rate, 
\begin{equation*}
    \textstyle P_{\mathrm{em}}(t)=\sum_{\tau = kT}^{t} S_\tau/[t - kT + 1],
\end{equation*}
and the expected transmission success rate without jamming, $P_{\mathrm{ex}}(t) = P_{s^{\mathrm{Tx}}_k,0}$.
To get a reliable estimate $P_{\mathrm{em}}(t)$, we wait for $t_\mathrm{w}$ time slots during which the belief remains unchanged. For $t\geq kT+t_\mathrm{w}$, the belief update is
\begin{equation}\label{eq:update_f}
    f_{k}^{\mathrm{Jm}}(t+1) = \mathrm{Proj}_{[\epsilon,1]}\{f^{\mathrm{Jm}}_{k,t} - \vartheta_{t-kT} [P_{\mathrm{em}}(t)-P_{\mathrm{ex}}(t)]\},
\end{equation}
where $\mathrm{Proj}_{[\epsilon,1]}\{\cdot\}=\min\{\max\{\cdot,1\},\epsilon\}$ with constant $\epsilon>0$ is close to zero, and $\vartheta_{t}=C_1 \left( 1 - e^{-C_2 t} \right)$ denotes the adaptive update rate with $C_1,C_2>0$. 

Now, the controller triggers the defense strategy under either of the two events: event $\cl{E}_{\mathrm{bad}}$ when $s_K^{\mathrm{Tx}}
=B$ and $f_{k}^{\mathrm{Jm}}(t) \geq \xi$ for a predefined threshold $\xi$; or event $\cl{E}_{\mathrm{insuff}}$ when $\mu_{s^{\mathrm{Tx}}_k}(L(t) +(k+1)T-t)<v$. 

\subsubsection{Controller channel state unknown at the controller}\label{sec:unknownchannel}
In this case, the controller computes beliefs about both the channel state and jamming presence based on transmission acknowledgments. Let the joint belief on the controller and actuator channels at the controller be
\begin{equation*}
    f_{k,q,d}(t) =  \mathrm{Pr}(s_k=(q,d)\mid \{S_{\tau}\}_{\tau=kT}^{t}),
\end{equation*}
with an initial value of $\pi^{\mathrm{Tx}}_q\pi^{\mathrm{Jm}}_d$. 
Then, the marginal belief on the controller channel is $f_{k,q}^{\mathrm{Tx}}(t)=f_{k,q,G}(t)+f_{k,q,B}(t)$. Beliefs $f_{k,q}^{\mathrm{Tx}}(t)$ and $f_{k}^{\mathrm{Jm}}(t)$ guide the adaptive selection of the control inputs per slot $\mu(k)$ and trigger the defense strategy.

The belief is updated as
\begin{equation*}
    f_{k,q,d}(t+1)   =\frac{f_{k,q,d}(t) \beta_{k,q,d,S_t}(t)}{\sum_{q',d'}f_{k,q',d'}(t) \beta_{k,q',d',S_t}(t)},
\end{equation*}
where we define
\begin{align*}
    \beta_{k,q,d,S}(t)
    &=\mathrm{Pr}(S_{t}=S|s_k=(q,d),\{S_{\tau}\}_{\tau=kT}^{t-1})\\
    &= 1-S-\sum_{a=0,1}\beta_{k,a}^{\mathrm{Jm}}(t) P_{q,d,a},
\end{align*}
using \eqref{eq:successprob}, where $\beta_{k,a}^{\mathrm{Jm}}(t)=\mathrm{Pr}(a_t=a|\{S_\tau\}_{\tau=kT}^{t-1})$ is the inferred jamming probability at time slot $t$. Since the jammer’s actions are based on an adaptive and feedback-driven strategy, it is hard to infer this probability. However, the jammer controller is inactive in the first slot $t=kT$ where $\beta_{k,a}^{\mathrm{Jm}}(t)=1$ for $a=0$. Therefore, at $t=kT$, the controller sends a pilot signal to the
actuator, and updates the belief about the controller channel. This step takes one time slot; control inputs are sent in the remaining $T-1$ slots. For $t>kT$, we conservatively set $\beta_{k,a}^{\mathrm{Jm}}(t)=0.5$ to get 
\begin{equation*}
    f_{k,q,d}(t+1)  \! =\!\frac{f_{k,q,d}(t) [1\!-\!S_t \!-\!(P_{q,0}\!+\!P_{q,d,1})/2]}{1\!-\!S_t\!-\!\sum_{q',d'}f_{k,q',d'}(t) [(P_{q',0}\!+\!P_{q',d',1})/2]}.
\end{equation*}
Then, the controller adaptively sets the control inputs per slot at time $t$ as $
    \mu(k)=\mu_{q^*}$ with $ q^*=\arg\max_q f_{k,q}^{\mathrm{Tx}}(kT+1)$.
    
While per-slot jamming actions are hard to infer, the controller estimates the likelihood of jamming within the block using the belief 
$f_{k}^{\mathrm{Jm}}(t)$ updated according to \eqref{eq:update_f} with $P_{\mathrm{ex}}(t) = \sum_{q \in \{G, B\}} f^{\mathrm{Tx}}_{k,q}(t) P_{q,0}$ using \eqref{eq:successprob2}.

Finally, the controller triggers the defense strategy under one of the two events: event $\cl{E}_{\mathrm{bad}}$ when 
\begin{equation*}
    f^{\mathrm{Tx}}_{k,B}(t) \geq \bar{\xi}\quad\text{and}\quad f_{k}^{\mathrm{Jm}}(t) \geq \xi,
\end{equation*}for predefined thresholds $\bar{\xi}$ and $\xi$; or event $\cl{E}_{\mathrm{insuff}}$ when \begin{equation*}
    \textstyle\left[\sum_{q\in\{G,B\}} f^{\mathrm{Tx}}_{k,q}(t)\mu_q\right]\!\left[\sum_{\tau=kT+1}^{t}\!S(\tau)\!+\!(k+1)T-t\right]
    \!<\! v.
\end{equation*} 

\vspace{0.1cm}

\section{Numerical Results and Discussion}
We present simulation results for our jamming scheme and compare it against three strategies: (i) random jamming without feedback (probability 0.5), (ii) constant jamming (every slot), and (iii) constant jamming with a limited budget (until exhaustion). We also evaluate our defense scheme performance. For simulations, we consider a generic dynamical system $(\bs{A},\bs{B})$ in \eqref{eq:controlsys_model} with a controllability index $v = 36$. We choose block length $T = 15$, and control inputs per slot $\mu_G=4$ and $\mu_B=3$. Rayleigh fading parameters are $\theta_G = 0.15$, $\theta_B = 0.65$, with SNR threshold $\gamma = 1$ and noise variances $\sigma_{\mathrm{Tx}}^2 = \sigma_{\mathrm{Jm}}^2 = 1$ and transmission power is $\eta = 1$. 

\begin{figure}[htbp]
        \centering
        \includegraphics[trim=72 26 88 25, clip,width=\linewidth]{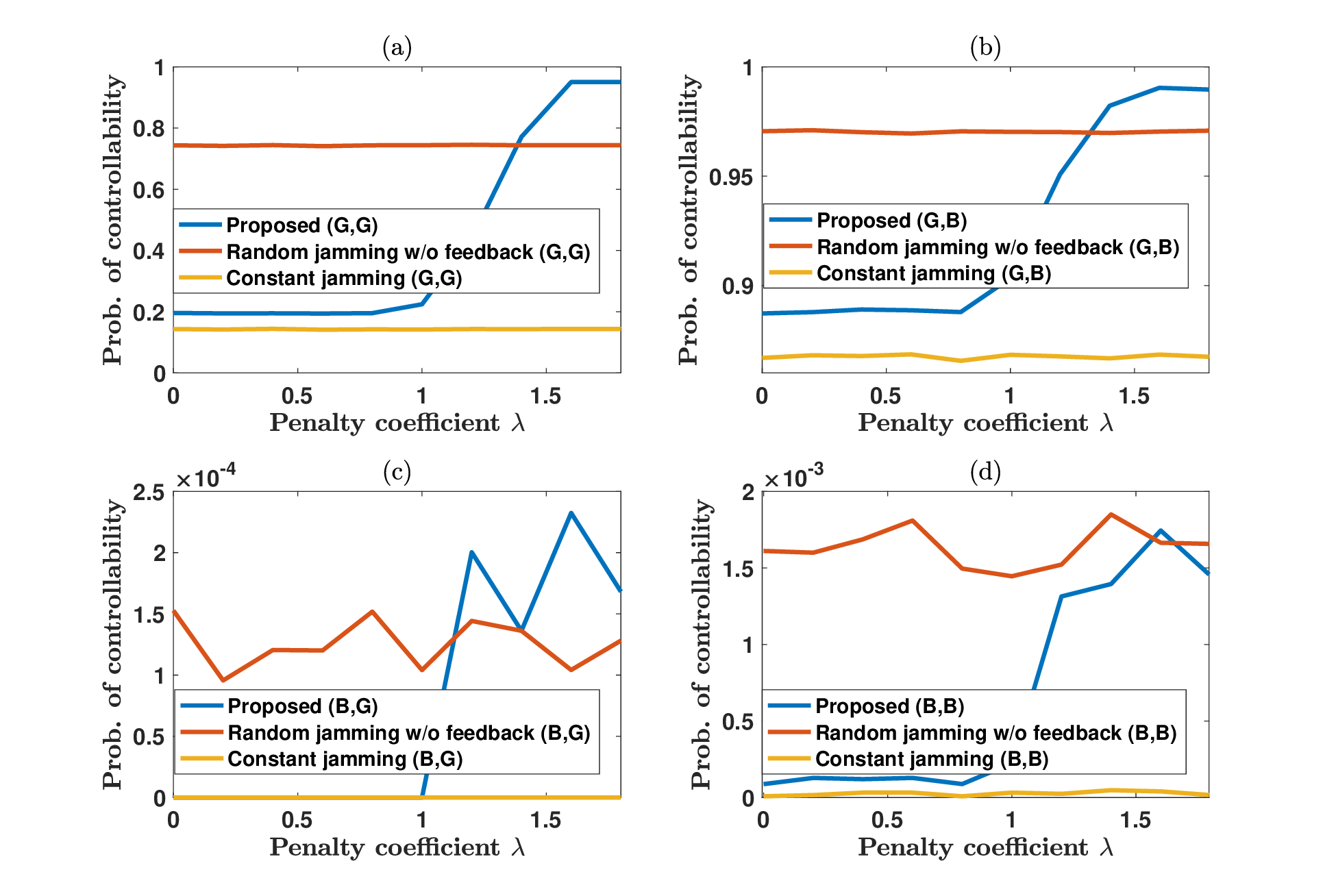}
        \caption{Probability of block controllability as a function of the penalty coefficient $\lambda$ for different channel states.
        }
        \label{fig:1}
\end{figure}
\begin{figure}[htbp]
        \centering
        \includegraphics[width =7cm,height=4.5cm]{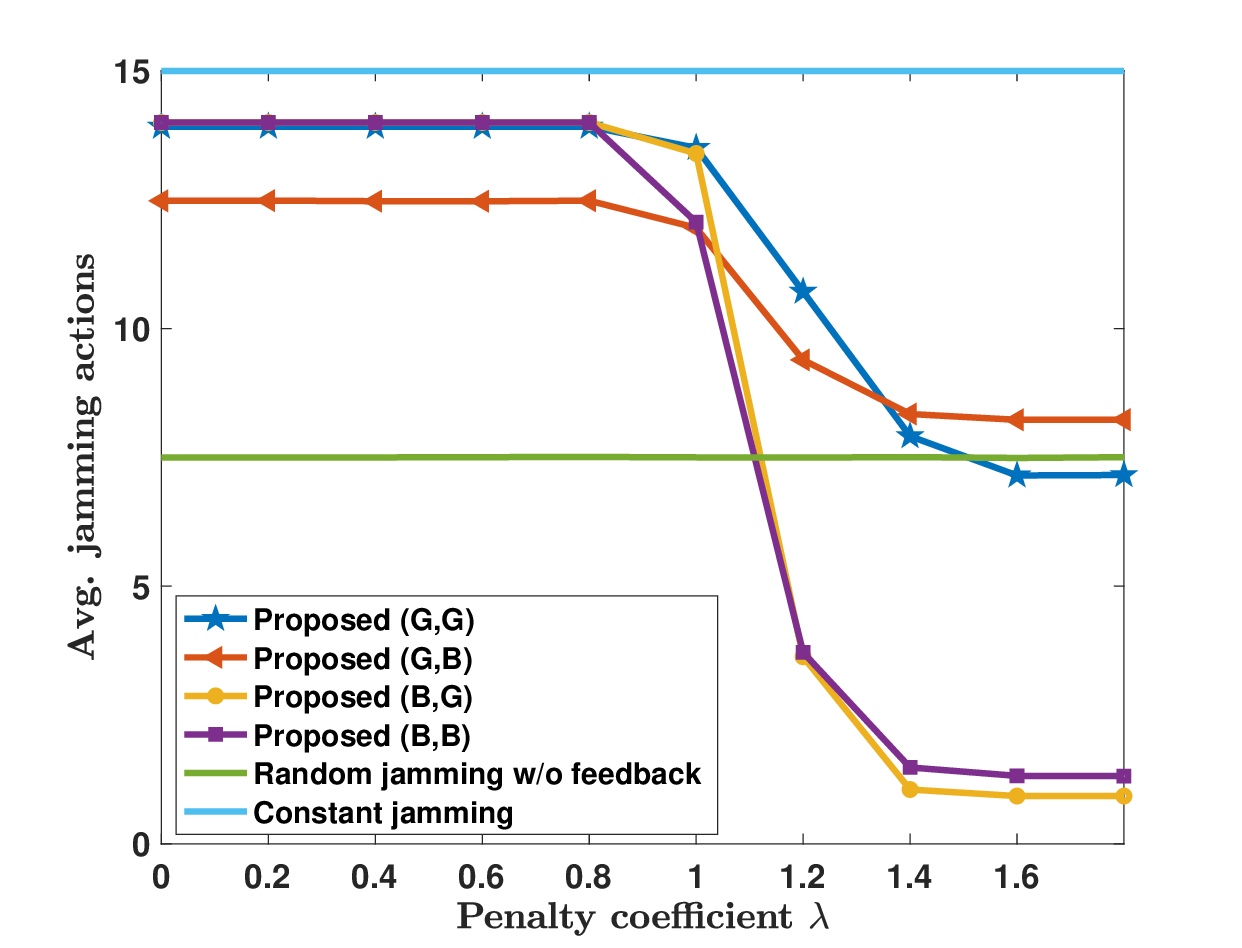}
        \caption{Average number of jamming actions as a function of the penalty coefficient $\lambda$ for different channel states.
        }
        \label{fig:2}
\end{figure}
\begin{figure*}[hptb]
	\centering
        \subfigure[]{
			\includegraphics[trim=0.8cm 0.2cm 0.8cm 0.68cm, width = 8cm,width=0.3\linewidth]{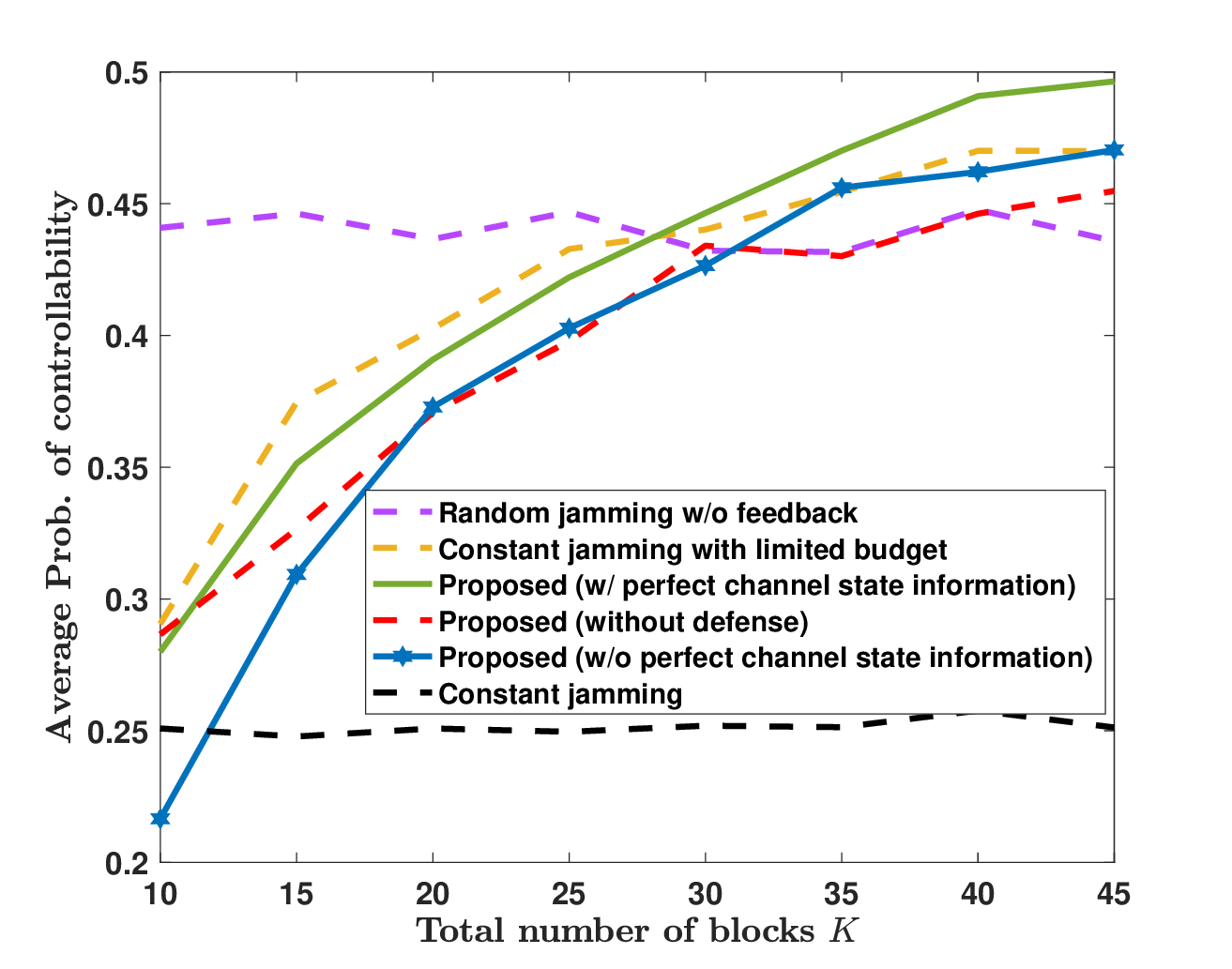}
		\label{fig:5}
	}
        \subfigure[]{
   		 	\includegraphics[trim=0.8cm 0.3cm 0.8cm 0.7cm,width=0.3\linewidth]{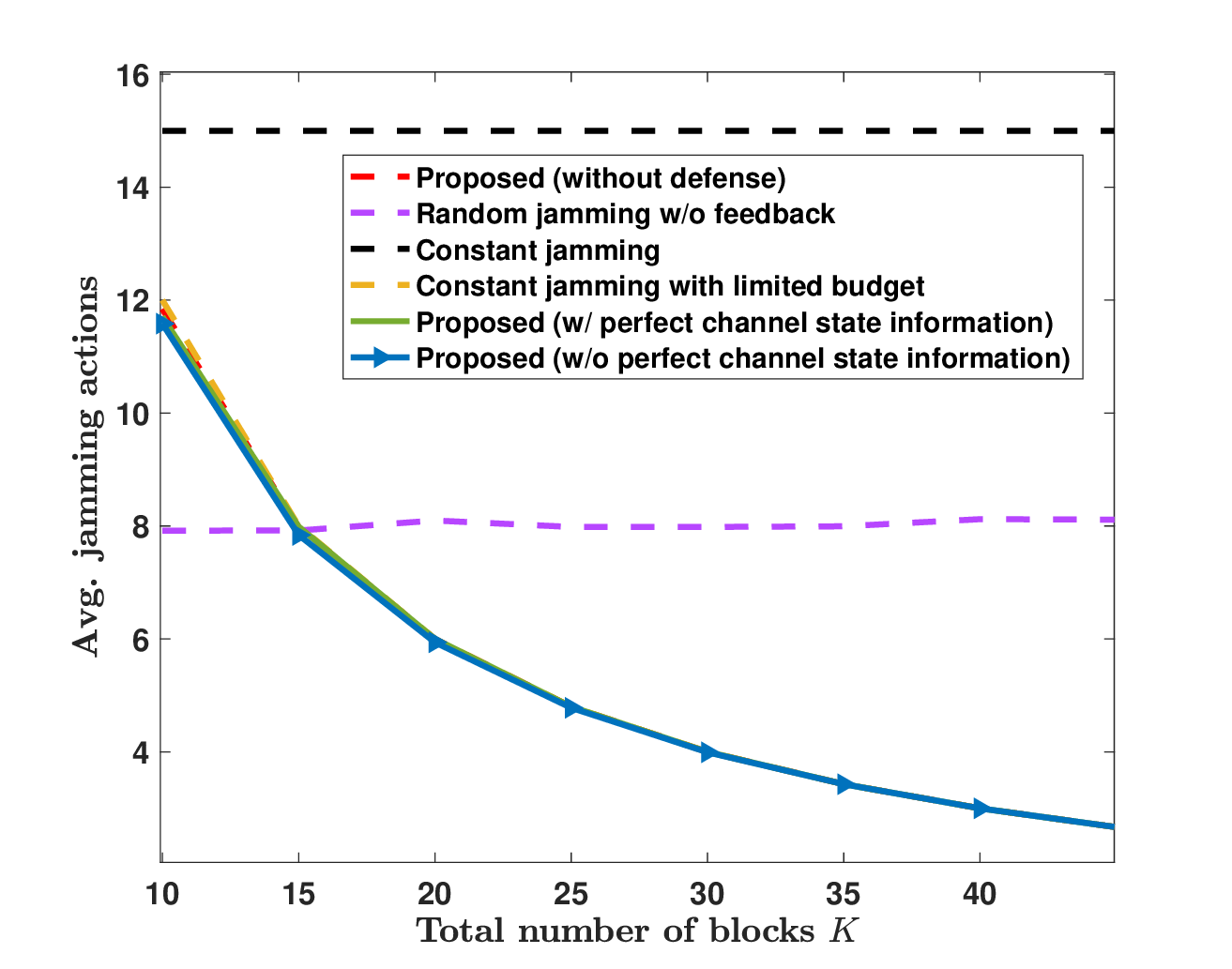}
	\label{fig:6}
        }
        \subfigure[]{
   		 	\includegraphics[trim=0.8cm 0.3cm 0.8cm 0.7cm,width=0.3\linewidth]{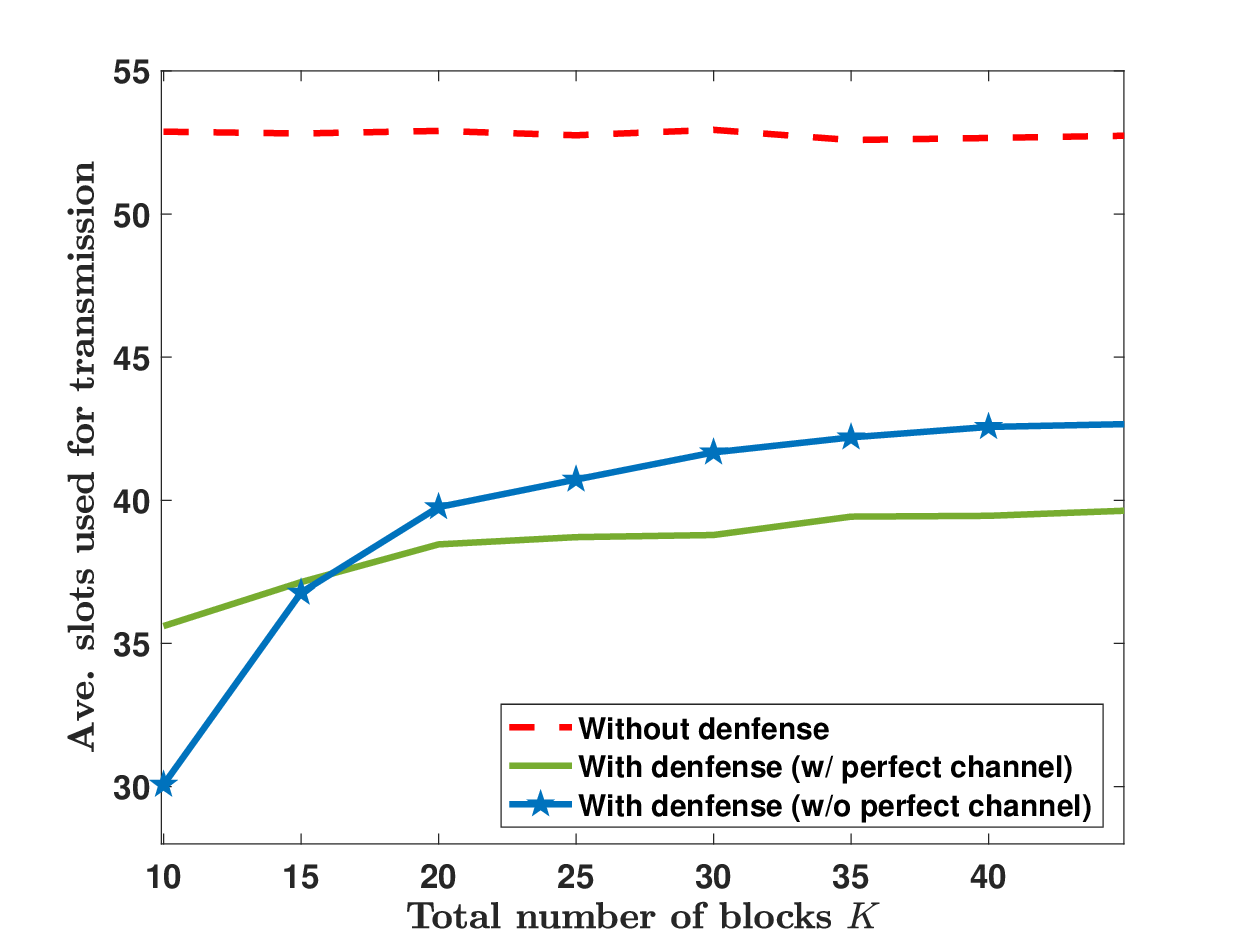}
	\label{fig: comparison_K2}
        }
	\caption{    Comparison of our jamming strategy, with and without defense, and other strategies: random jamming, constant jamming, and limited-budget constant jamming, across varying block numbers and averaged across different channel states.}
	\label{fig: comparison_K}
\end{figure*}
Figs.~\ref{fig:1} and \ref{fig:2} show the impact of the penalty coefficient $\lambda$ on the block controllability and jamming actions for single-block optimization, assuming that the controller knows its channel. Here, initial channel state distributions are uniform: $\pi^{\mathrm{Tx}}_q=\pi^{\mathrm{Jm}}_d=0.5$ for $q,d\in\{G,B\}$. For lower values of $\lambda$, our jamming strategy can effectively degrade the block controllability. As $\lambda$ increases, its impact decreases due to higher penalties. Still, for the poor controller channel, i.e., $(B,G)$ and $(B,B)$, the strategy consistently performs well across all $\lambda \in [0,1.8]$. 
Fig.~\ref{fig:2} shows that, as $\lambda$ increases beyond 1, the strategy uses fewer jamming actions, conserving energy. This effect is most pronounced in $(B,G)$ and $(B,B)$ states, where poor controller channels make jamming more effective. Thus, proper tuning of $\lambda$ enables the jammer to achieve effective jamming performance using limited resources. Our jamming strategy does not achieve the same level of control performance degradation as constant jamming. Yet, it significantly reduces the average number of jamming actions while maintaining an acceptable level of performance degradation with an appropriate penalty $\lambda$ (e.g., $\lambda=1.2$). 
The random jamming strategy performs the worst in terms of both controllability degradation and energy efficiency.

Fig.~\ref{fig: comparison_K} shows the control performance as a function of the number of blocks $K$ in the multi-block scenario under different jamming strategies, as well as a comparison between systems with and without the defense scheme. Here, we set $\pi^{\mathrm{Tx}}_G = \pi^{\mathrm{Jm}}_B=0.3$ and $\pi^{\mathrm{Tx}}_B=\pi^{\mathrm{Jm}}_G =0.7$, with transition probabilities $p_{G,G} = p_{B,B} = 0.7$, and $p_{G,B} = p_{B,G} = 0.3$, for both channels. The jamming budget is $M = 8T$.
The defense scheme parameters are set to $\epsilon=0.01$, $\bar{\xi}=0.8$, $\xi=0.6$, $\vartheta=0.1$ and $t_\mathrm{w}=3$. The penalty coefficients are $\lambda_{G,G}=0$, $\lambda_{G,B}=0.7$, and $\lambda_{B,G}=\lambda_{B,B}=0.8$.
 
Fig.~\ref{fig:5} shows that the defense scheme with perfect controller channel state knowledge outperforms the one without it. Initially, the scheme without perfect controller channel state has much lower block controllability probability than the {\it without defense} case, since $\mu(k)$ is chosen only based on belief, while the {\it without defense} case assumes perfect channel state information, resulting in an unfair comparison. However, as $K$ increases, both defense schemes improve block controllability probability, demonstrating long-term effectiveness under adaptive jamming. From Fig. \ref{fig:6}, the constant jamming consumes the maximum budget in every block, thus achieving the greatest block controllability degradation. In contrast, the proposed jamming strategy without defense significantly reduces the number of jamming actions as $K$ increases, while still maintaining a competitive level of block controllability degradation. Moreover, it achieves a low average probability of block controllability when $K$ is relatively small, outperforming both random jamming and constant jamming with a limited budget. However, as $K$ increases, the block controllability degradation becomes less pronounced, as the budget is spread over more blocks, reducing its impact per block. The random jamming strategy again yields the worst performance. Finally, Fig.~\ref{fig: comparison_K2} shows that the average number of communication slots used by the controller is significantly reduced with defense. This saving is greater with perfect controller channel state, particularly as $K$ grows.

\section{Conclusion}
We studied resilient control in rested linear systems under adaptive jamming attack. We considered a worst-case scenario where the jammer adaptively optimizes its strategy based on a joint belief of its own channel state and that of the controller-actuator channel, under a limited attack budget. To counteract adversarial behavior, we presented an event-triggered defense scheme guided by belief updates of the controller channel state. Simulation results showed that this scheme enhanced system resilience while reducing the communication overhead. Future work can extend this to unreliable sensor-controller channels (with or without jamming), multi-sensor setups.

\bibliography{references.bib}
\bibliographystyle{IEEEtran}

\end{document}